\documentclass[12pt]{article} 
\usepackage{amsmath}
\usepackage{graphicx}
\begin{document}

% Title of the article
\title{Description of conductivity steps in polymer and other materials by functions of p-adic argument}

% Authors
\author{Viktor Zharkov \\
Natural Science Institute of Perm State University\\
  Genkel st.4, Perm, Russia, vita@psu.ru }
\maketitle   % please do not remove

\begin{abstract}
We study the quantum jumps of physical quantities in a strongly correlated many  electron systems based on a new p - adic functional integral approach. It is shown that a description in terms of the p - adic numbers leads to the fractal behavior and can describe the quantum jumps in the conductivity and magnetization as a function of voltage and magnetic field in nanotechnology devices and low-dimensional strongly correlated organic metals and other materials.
\end{abstract}

\section{P-adic numbers in description of conductivity steps}

•	Many low-dimensional materials exhibit jumps of conductivity and magnetization when the applied voltage and magnetic field are changed \cite{c1,c2,c3,c4}. 
•	
•	 Usually the very particular mechanism is involved for the explanation of these phenomena.
•	I would like to emphasize that in this area it is desirable to have a general approach that explains these quantum phenomena through the study of the nature of strong electron's correlations which are usually present in these compounds.
•	
•	In this paper I demonstrate that the jumps in the conductance as a function of voltage in the polymer and other material  can be described by functions on p-adic argument. Let's give some figures that demonstrate those properties which we want to describe. In figure 1 \cite{c1} equidistant steps clearly visible which form the basic fractal level. In the inset  secondary structure of the fractal is clearly visible.
Note that in our approach, a large number of steps of the same height can be obtained by taking a large prime number p, which specifies the fractal. In figure 2 \cite{c2}  one can see Shapiro steps observed in a dc superconducting quantum interference device. It is seen that this steps have also two level structure. 
On this chart two systems with different heights of the steps are  clearly visible. These two type of steps form a system of two levels of the fractal description. The  p-adic function which I  proposed  qualitatively exactly reflects this pattern.

In figure 3 \cite{c3} fractal structure of two types  is visible: descending steps of varying heights and saw, which contains at least three levels of nested saw with different height of jumps. 

In figure 4 it is obviously seen that we  need to have a different number of levels depending on the accuracy of the fractal description \cite{c4} . We have here poly(para-phenylene ethynylene)  (PPE) and derivative of PPE with thioacetate end group (TA-PPE). In first insert  aggregate is shown in the form of steps. On the second insert we have the saw. Our problem is  to find such a p-adic function, which gives this behavior

\begin{figure}[ht]
\centering
\includegraphics*[width=200pt,height=121pt]{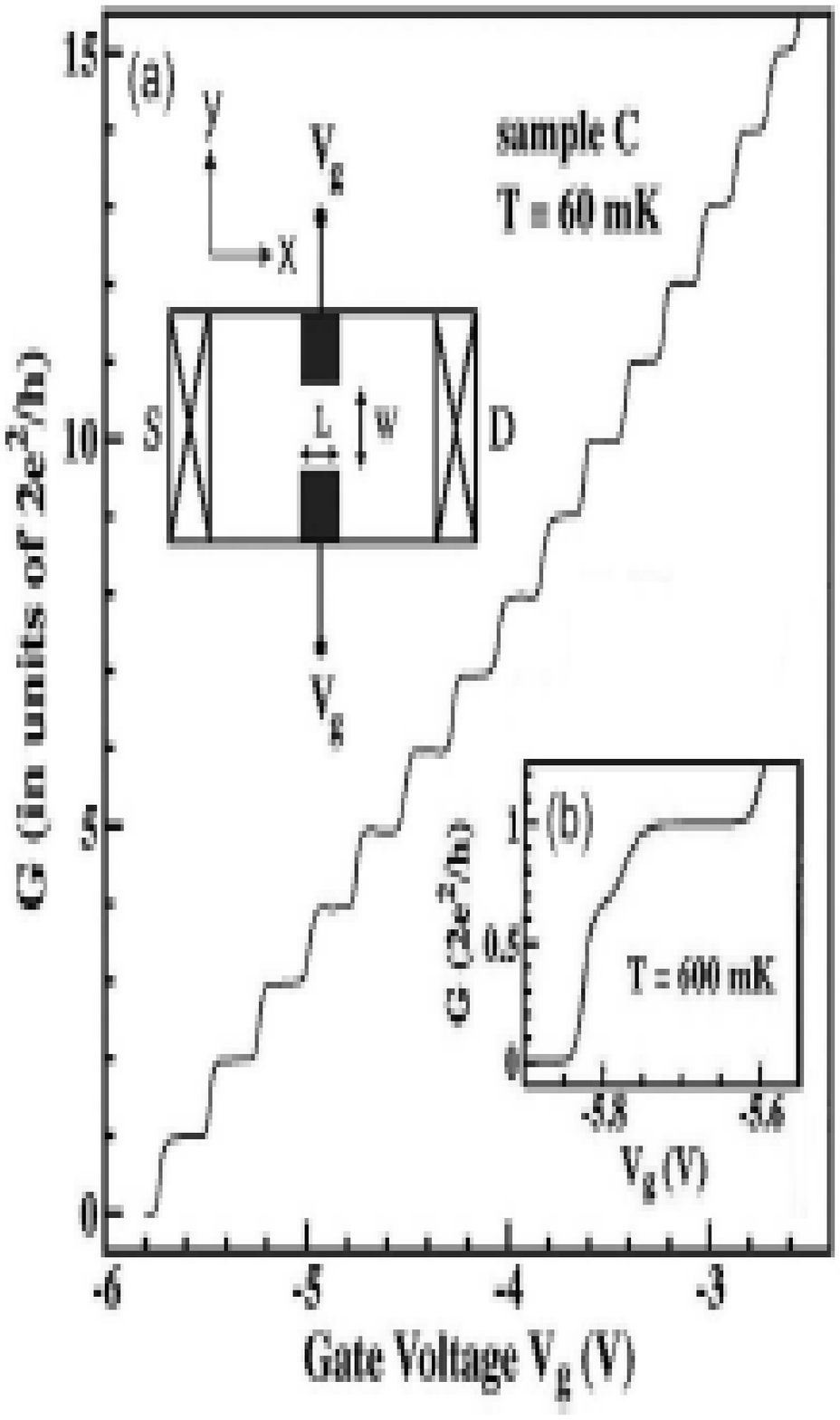}
\caption{The linear differential conductance vs. voltage V from quantum point contact.}\label{fig:name1}
\end{figure}

\begin{figure}[ht]
\centering
\includegraphics*[width=200pt,height=121pt]{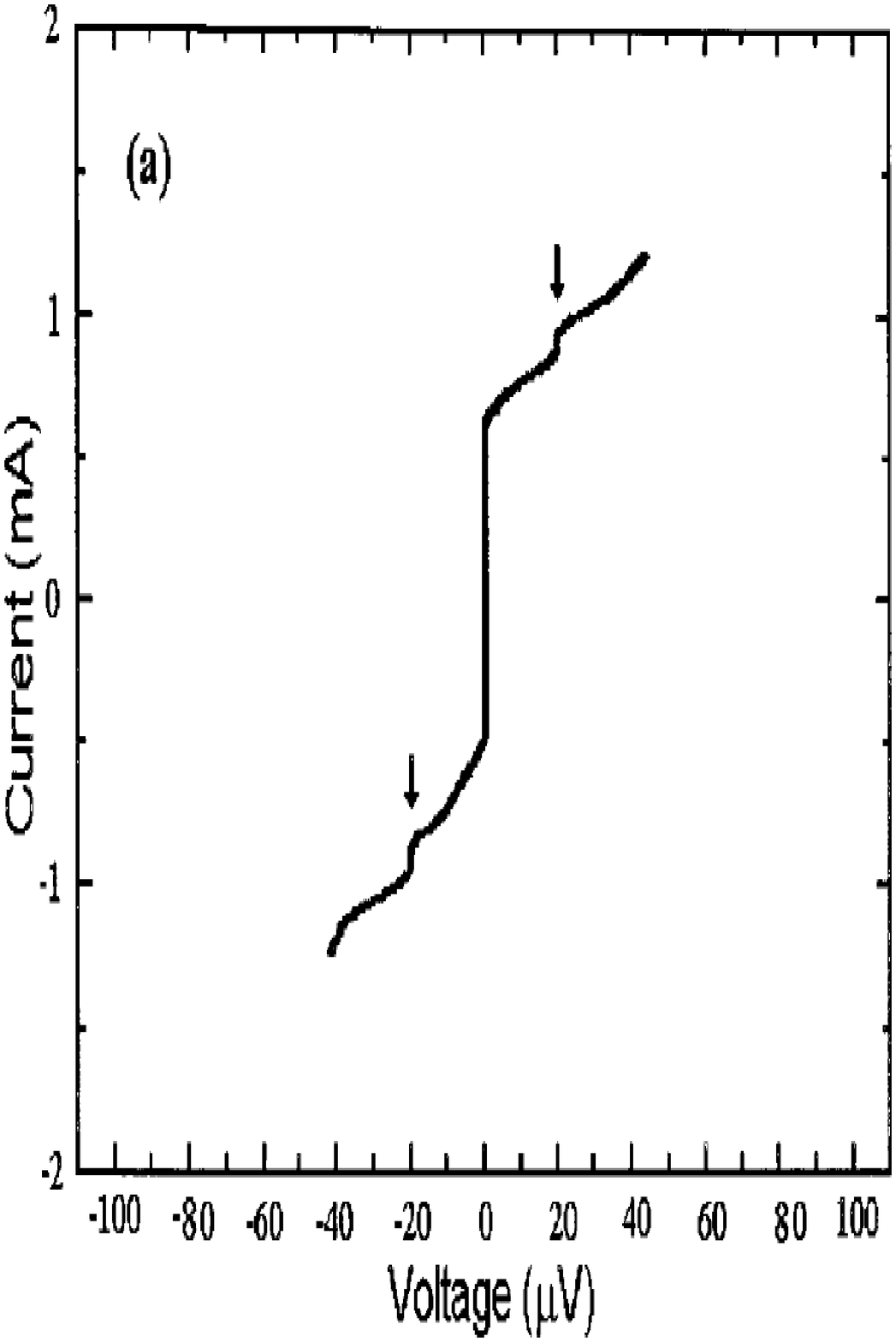}
\caption{Rf power dependence of the Shapiro steps on the I-V curve of the dc SQUID. First and second Shapiro step with normal separation.}\label{fig:name2}
\end{figure} 
 
\begin{figure}[ht]
\centering
\includegraphics*[width=200pt,height=121pt]{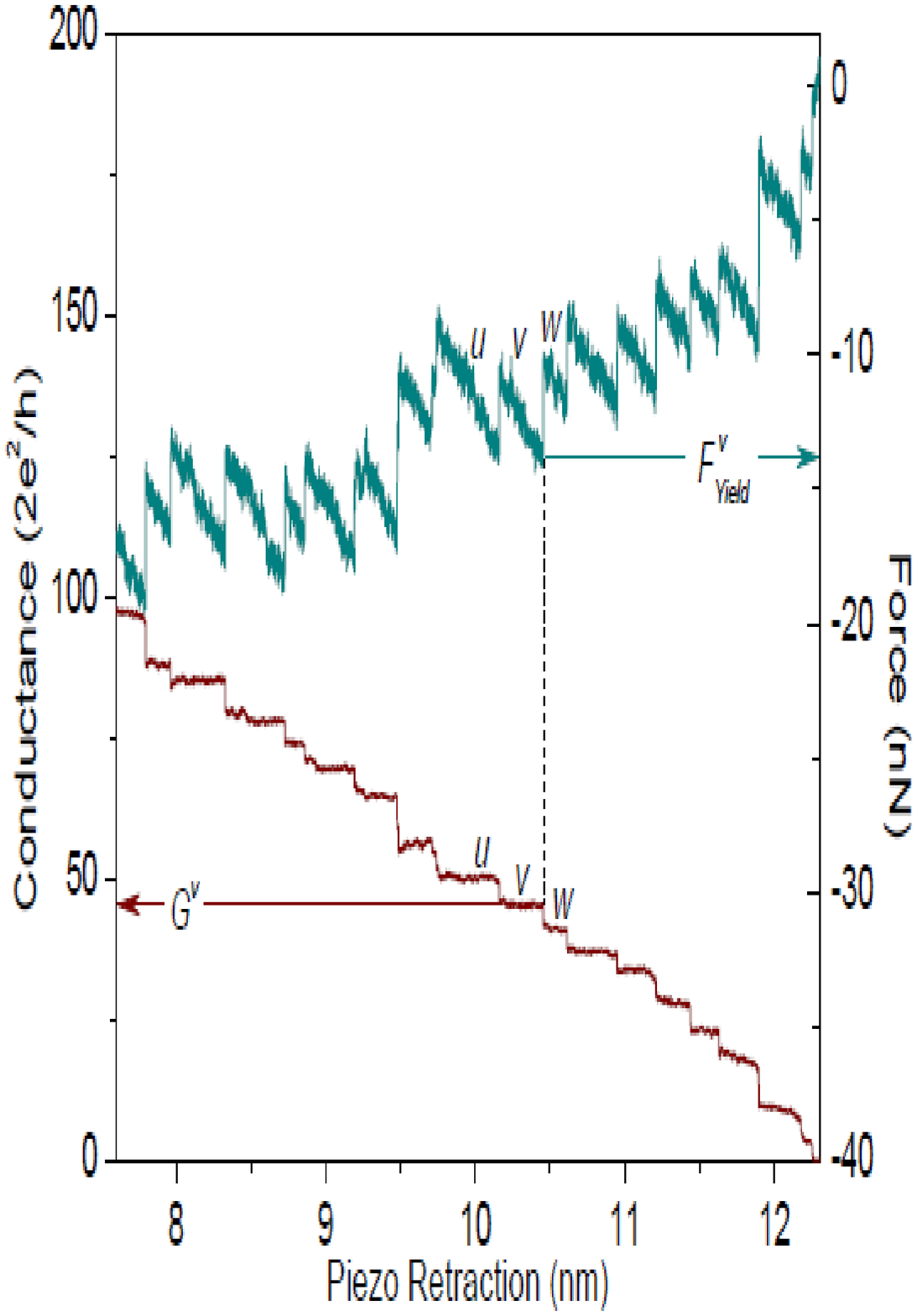}
\caption{Simultaneously measured force and conductance during deformation of atomic-sized bridges in different conductance regimes.}\label{fig:name3}
\end{figure}

\begin{figure}[ht]
\centering
\includegraphics*[width=200pt,height=121pt]{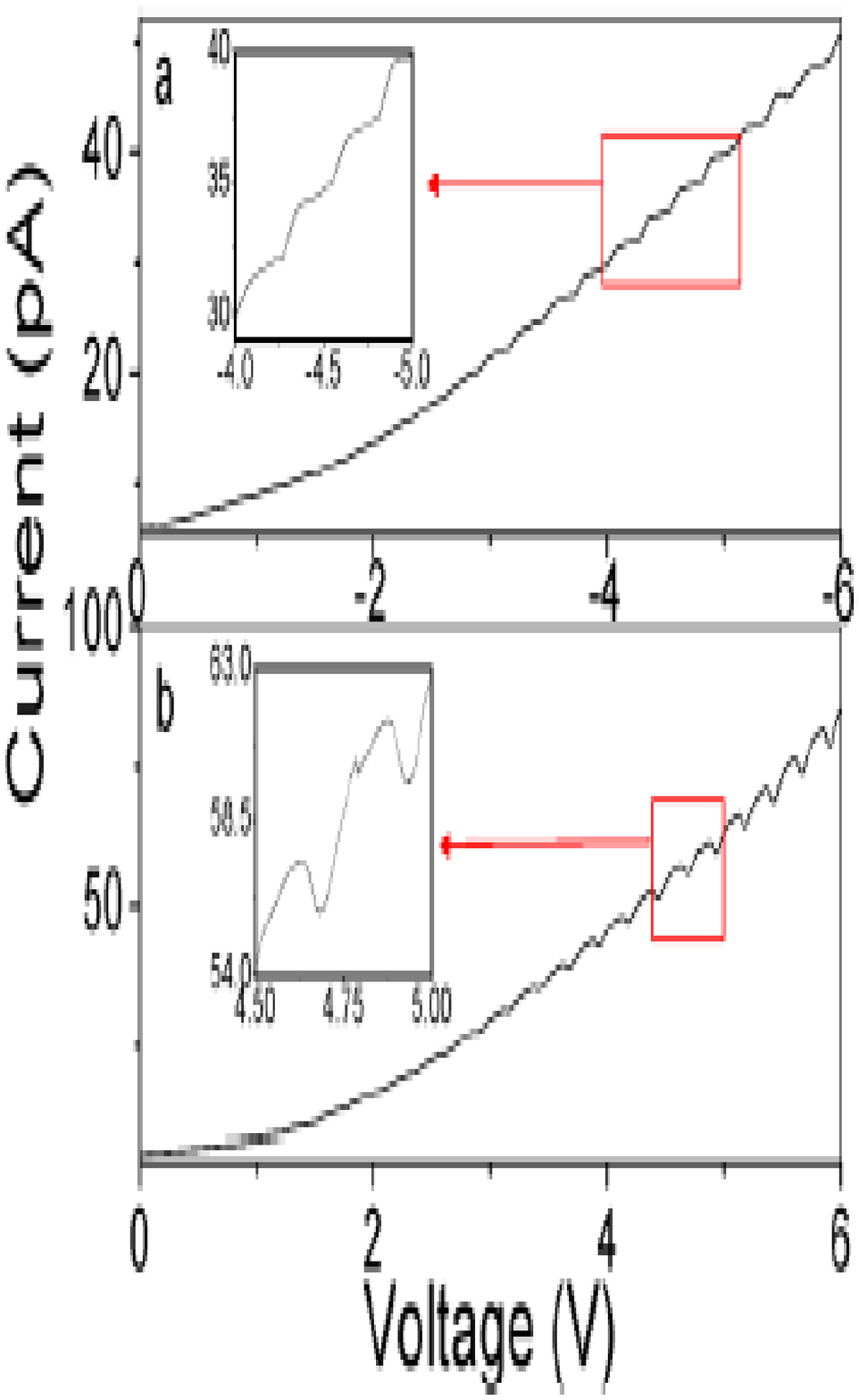}
\caption{I-V characteristics of a TA-PPE molecular junction. a) Negative scan direction: inset, enlarged steps from -4.0 to -5.0 V. b) Positive scan direction: inset, enlarged step from 4.5 to 5.0 V.}\label{fig:name4}
\end{figure}

\section{ P-adic numbers and functions}

Let's briefly give  the necessary information about the p-adic numbers and functions on this numbers.

Let us define some basic notation. P-denote a prime number. An
arbitrary rational number $ r $ can be written in the form
$r=p^{\nu }\frac{m}{n}$ with n and m not divisible by p. The
p-adic norm of
the rational number is equal to $\mid r\mid _{p}=p^{-\nu }$, $%
\mid 0\mid _{0}=0$. The field of p-adic numbers $ Q_p$ is the
completion of the field of rational numbers $ Q $ with  the p-adic
norm . 

Let us remind that a real number may be expressed by the following expansion $r={\sum}_{-\infty }^{N}a_{k}p^{k};\quad
a_{k}=(0,1,....,p-1);k\in Z.$

Integer P-adic numbers may be expressed  by the following expansion $r={\sum}^{N }_{0}a_{k}p^{k};\quad
a_{k}=(0,1,....,p-1);k\in Z.$

We construct the following function for description of fractal properties which are described in our figures: $ f_b(r)={\sum}_{0}^{N }$ $a_{k}p^{bk};\quad
a_{k}=(0,1,....,p-1);k\in Z.$  Here parameter b defines the fractal dimention.
We compute $ f_b(r)$ and make the plots (fig. 5, fig. 6)  in coordinates $ r,f $ for different value of b.  

Computer simulation of the function $ f_b(r) $ for different values of the  fractal dimensions is shown in fig.5.6

The argument on this graph can be associated with a voltage V and the function  gives us the conductance.

 It is shown that this function  contains two types of fractal behavior, shown in Figure 4.

\begin{figure}[ht]% 
\centering
\includegraphics*[width= 200pt,height=121pt]{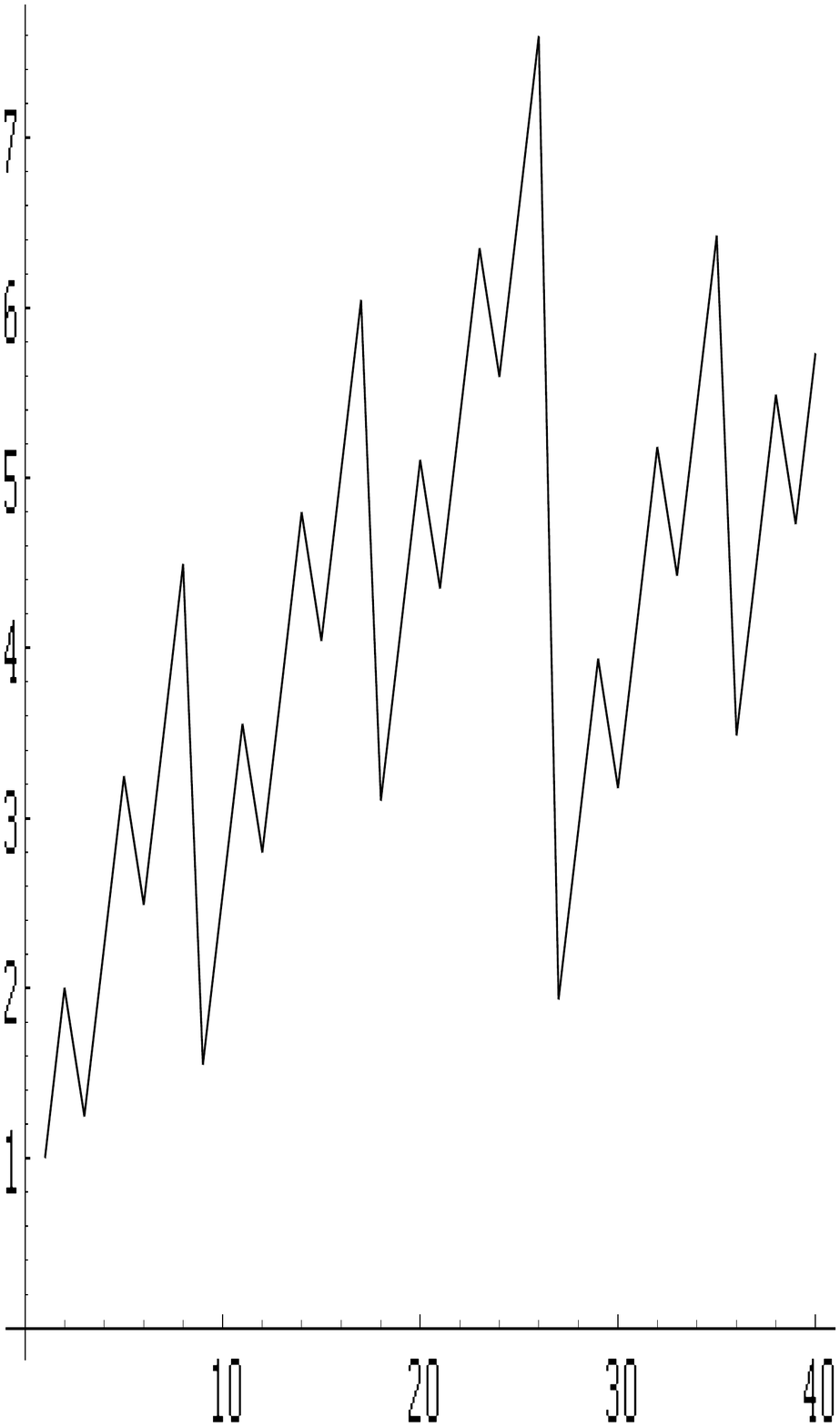}
\caption{Subcritical wave with $ b<1$.}
\label{onecolumnfigure}
\end{figure}

\begin{figure}[ht]
\centering
\includegraphics*[width=200pt,height=121pt]{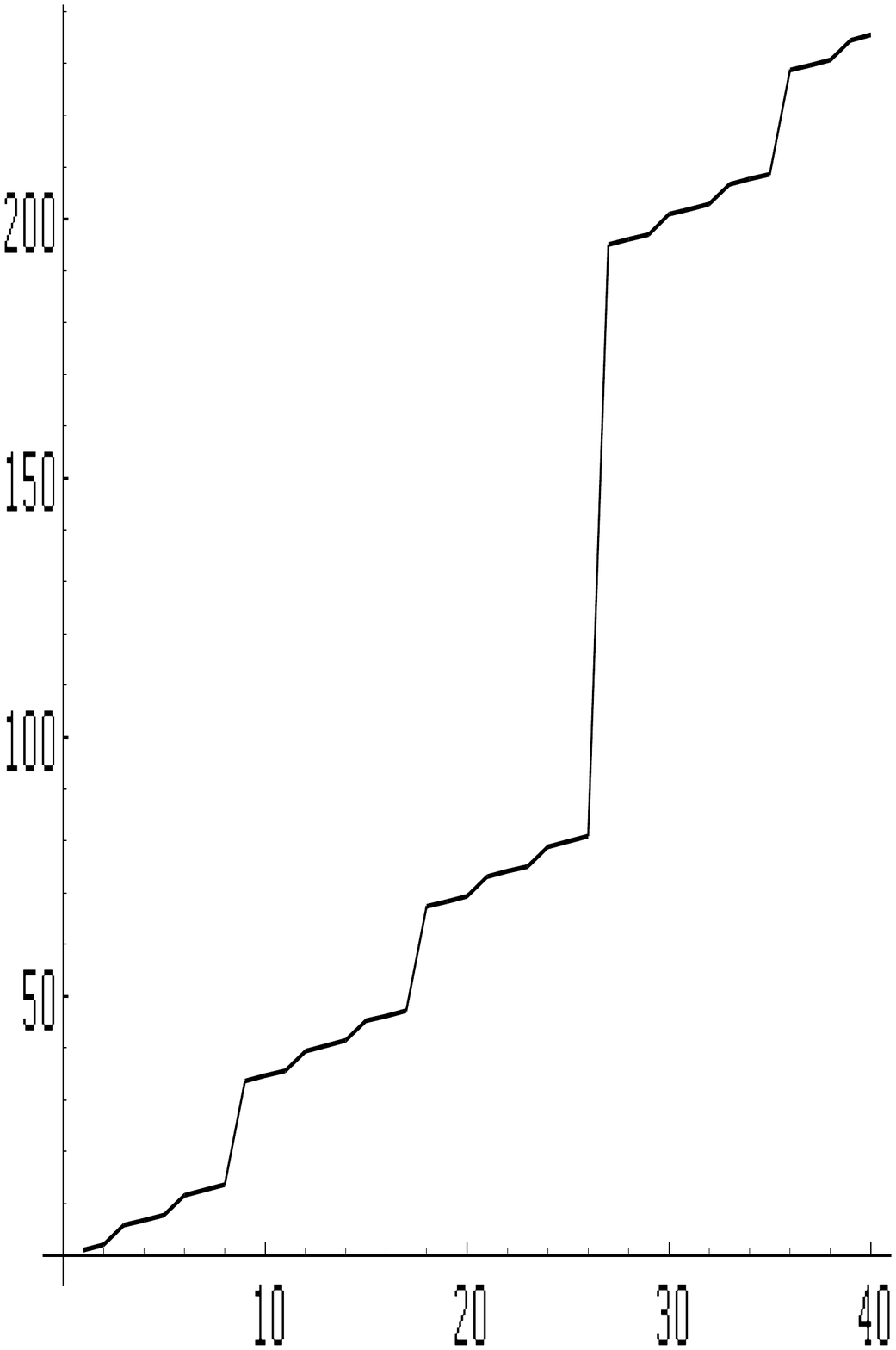}
\caption{Supercritical wave for $ b>1 $}\label{fig:name5}
\end{figure} 

\section{Physical derivation of p-adic description}

In this paper we  give examples of the functions of the p-adic arguments to describe the conductivity jumps.
It was noted that in the experiment in jumps the fractal structure is visible and have the same property as the function of the p-adic arguments . Jumps and fractal behavior appears in the low-dimensional systems, and our view is that the use of p-adic numbers is promising in this area. Let us consider the physical meaning of using the p-adic numbers to describe the fractal behavior of various quantities. In  paper \cite{b6} it was pointed out that the p-adic numbers are very well optimized for the description of fractals. For example, a field of 2-adic numbers is the Cantor set.
The overall picture that can be given on the basis of the author's papers  is as follows.
As a rule, low-dimensional systems are in the strong correlation regime.
In \cite{b2,b3}  a novel approach to many-electron systems based on nonlinear functional integral has been developed.
The presence of nonlinear terms leads to the induction in the system  of some quantum symmetry given by the Hopf algebra when strong Coulomb repulsion exist in system. 
For example, the presence in a system of quantum symmetry given by the group $ SU_q (2) $ leads to deformation of the integration measure  to the Jackson's integral.

It is known \cite{b6} that when $ q=1/p $ we arrive to p-adic functional integral.
And as already noted, in the p-adic approach  fractals appear very natural.
That is why the strong electron correlations arise fractal behavior of the conductivity and magnetization.
Strict appearance of measures of Jackson's type in the Hubbard model is shown by the author.

In conclusion, we showed that  the proposed p-adic function qualitatively well describes the two type of conduction as a function of voltage in polymer TA-PPE.

Let us briefly give a general scheme for obtaining the effective functional for the Hubbard model,
proposed in the papers \cite{b2,b3} and further developed in recent papers .
We start from the Hubbard model, written in the usual creation and annihilation operators:

\begin{equation} H=-W\sum_{ij\sigma }\alpha _{\sigma ,i}^{+}\alpha _{\sigma ,j}+U\sum_{i,\sigma \
}n_{\sigma ,i}n_{-\sigma ,i}+\mu \sum_{\sigma ,i}n_{\sigma ,i}, \label{h1} \end{equation}

here $ \alpha_ {\sigma, i }^{+}, \alpha_ {\sigma, j} $ -  creation and annihilation operators, $ n_ {\sigma
, i} $-the electron density operators, $ W, U, \mu $-width of the conduction band, single-site
repulsion of two electrons and chemical potential.

Second term is diagonal in the following "atomic" basis:

$$|0\succ ;|+\succ =\alpha_{\uparrow }^{+}|0\succ; |-\succ =\alpha_{\downarrow }^{+}|0\succ;|2\succ
=\alpha_{\uparrow }^{+}\alpha_{\downarrow }^{+}|0\succ . $$

All operators in this basis can be expressed in 16 of Hubbard operators, and can
be divided into fermions and bosons. In terms of these operators  the Hubbard model takes the following form:

\begin{equation} H=U\sum_{i,p}X_{i}^{pp}-W\sum_{ij\alpha \beta }X_{i}^{-\alpha }X_{j}^{\beta }
\label{HubbAtomic} \end{equation}

Next we  use a functional formulation of many-electron systems, proposed in
work \cite {b2}.
According to this approach, the evolution operator between initial and final states is given by the following functional integral with action, expressed through
effective functional, which is calculated (\ref{HubbAtomic}) using supercoherent state.

\begin{equation} <G_{f}|e^{-iH(t_{f}-t_{i})}|G_{i}>=%
\int_{|G_{i}>}^{|G_{f}>}D(G,G^{\ast })e^{-iS[G,G^{\ast }]}; \label{fi}
\end{equation}

here the action is the following expression:

\begin{equation}
 S[G,G^{\ast
}]=\int_{t_{i}}^{t_{f}}dt\int_{V}d^3r\frac{<G(r,t)|i\frac{\partial }{\partial
t}-H|G(r,t>}{<G(r,t|G(r,t>} \label{lag}
\end{equation}

We describe the local properties of strongly correlated systems by following function, which is 
equal to the supercoherent state:

\begin{equation}
\mid G>=\exp \left( \begin{array}{cccc} E_{z} & 0 & 0 & E^{+} \\ \chi _{1} & H_{z} & H^{+} & 0 \\
\chi _{2} & H^{-} & -H_{z} & 0 \\ E^{-} & -\chi _{3} & \chi _{4} & -E_{z}\end{array}\right) \mid
0> \label{Scs}
\end{equation}

In the exponent we have the dynamic fields that depend on the coordinates and time. The electric field is given by the  three-dimensional vector. The magnetic field has three components, depending on the spatial and temporal coordinates.
Fermionic fields are given by odd grassmann's valued functions of the coordinates and time.

% Use the following code if you wish to generate your bibliography with BibTeX;
% replace the string "pss-demo" below with the name(s) of
% the BibTeX data base(s) you want to use.
% The resulting bibliography-output (the content of the .bbl file)
% must be pasted back into this file before submission.
% Please also include your BibTeX data base file(s) in your submission
% so that we can re-run BibTeX if necessary.
%
%\bibliographystyle{pss}
%\bibliography{pss-demo}
%
% Replace the following example bibliography with your references
% before submission:

\end{document}